\documentclass[prb,floatfix,twocolumn,showpacs,amsmath,amssymb]{revtex4}
\usepackage{graphicx}
\usepackage{dcolumn}
\usepackage{bm}

\begin{document}

\title{Backflow correlations in the Hubbard model: an efficient tool for the 
metal-insulator transition and the large-$U$ limit}
\author{Luca F. Tocchio,$^{1}$ Federico Becca,$^{2}$
        and Claudius Gros$^{1}$ 
        }
\affiliation{
$^{1}$ Institute for Theoretical Physics, 
       Frankfurt University, 
       Max-von-Laue-Stra{\ss}e 1, D-60438 Frankfurt a.M., Germany \\
$^{2}$ CNR-IOM-Democritos National Simulation Centre 
       and International School for Advanced Studies (SISSA), 
       Via Bonomea 265, I-34136, Trieste, Italy
            }

\date{\today} 

\begin{abstract}
We show that backflow correlations in the variational wave function for the
Hubbard model greatly improve the previous results given by the Slater-Jastrow
state, usually considered in this context. We provide evidence that, within
this approach, it is possible to have a satisfactory connection with the 
strong-coupling regime. Moreover, we show that, for the Hubbard model on the 
lattice, backflow correlations are essentially short range, inducing an 
effective attraction between empty (holons) and doubly occupied sites 
(doublons). In presence of frustration, we report the evidence that the metal 
to Mott-insulator transition is marked by a discontinuity of the double 
occupancy, together with a similar discontinuity of the kinetic term that does
not change the number of holons and doublons, while the other kinetic terms are
continuous across the transition. Finally, we show the estimation of the charge
gap, obtained by particle-hole excitations {\it \`a la Feynman} over the 
ground-state wave function.
\end{abstract}


\maketitle

\section{Introduction}\label{sec:intro}

Backflow correlations were introduced by Feynman and Cohen~\cite{feynman} in 
order to obtain an accurate description of the excitation spectrum of liquid 
He$^4$. Indeed, by accounting for these correlations, it was possible to
have a marked improvement in the one-phonon dispersion curve and obtain an
accurate description of the roton excitations. Indeed, it was realized
that a picture of independent elementary excitations had to be extended to
account for their interactions. Essentially, the current associated with a
Feynman excitation involves a contribution from the backflow of the fluid
around it. Feynman and Cohen showed that, in the simplest case of a particle
tearing through the liquid at a given velocity, the pattern of induced 
longitudinal current far away from the particle has a dipolar 
form.~\cite{feynman} Then, the concept of backflow has been extended to 
weakly-correlated electron systems, where it turned out to be crucial for 
improving the description of the electron gas in two and three dimensions, 
in particular, for correlation energies and pair distribution 
functions.~\cite{ceperley} Further applications concerned metallic 
Hydrogen~\cite{holzmann} and small atoms or molecules,~\cite{molecules} where 
significant improvements in the total energy have been obtained.
  
Very recently, backflow correlations have been successfully applied to 
strongly-correlated electron systems, such as the Hubbard model and its
generalizations to include frustrating terms, improving the variational wave 
functions used so far to approximate the exact ground state. In particular,
in presence of frustration (e.g., for the $t{-}t^\prime$ Hubbard model on the 
square~\cite{tocchio1} and the triangular lattices~\cite{tocchio2}) 
a spin-liquid phase has been stabilized only thanks to backflow correlations.
This new term allowed us also to observe a renormalization of the underlying 
Fermi surface to perfect nesting at the metal-insulator 
transition.~\cite{tocchio3} Backflow correlations have been also recently 
applied to the infinite-$U$ Hubbard model to study the stability of the Nagaoka
ferromagnetism.~\cite{carleo}

When compared to former variational wave functions for frustrated Hubbard 
models,~\cite{fazekas} (short-range) backflow correlations significantly 
improved the description of ground-state properties with respect to 
density-density Jastrow factors~\cite{capello} and holon-doublon binding 
factors.~\cite{yoko} However, also in presence of backflow terms, the 
long-range Jastrow factor is still a fundamental ingredient of the wave 
function in order to describe the metal-insulator transition occurring at a 
finite value of the ratio between the electron-electron repulsion $U$ and the 
hopping integral $t$.

In this paper, we show the accuracy of backflow correlations and compare 
them with the $S$-matrix (i.e., strong-coupling) 
expansion.~\cite{macdonald,gros87} 
The latter approach becomes exact for $U/t \to \infty$ and define an iterative
procedure which results in an expansion in powers of $t/U$. As a consequence,
the (exact) wave function will be an admixture of electronic configurations 
with weights proportional to $(t/U)^n$, where $n$ is the number of 
doubly-occupied sites. Unfortunately, dealing with such a state is a 
difficult task, and, at present, only few attempts have been pursued in this 
direction.~\cite{baeriswyl} On the contrary, our approach based upon backflow 
correlations define a many-body wave function that can be easily treated by 
standard Monte Carlo methods. Most importantly, it remains accurate also for
intermediate electron-electron repulsion, allowing a precise description of 
the physics down to the metal-insulator transition.~\cite{tocchio1,tocchio2} 
Here, we show that the energy gain due to backflow correlations in the 
insulating phase is mainly related to a better description of the local hopping
processes changing the total numbers of holon-doublon couples. We discuss the 
extension of backflow correlations to long-range distances and how the behavior
of the Jastrow factor is affected by the presence of backflow correlations. 
We present the results both for one and two dimensions, with the variational 
wave function containing no magnetic terms. Finally, we discuss also how the 
variational method allows us to calculate the charge gap in the insulating 
phase and present results for the one-dimensional case. 

The paper is organized as follows: in Sec.~\ref{sec:model}, we introduce the 
Hamiltonian and we describe our variational wave function; 
in Sec.~\ref{sec:Smatrix}, we give a short review of the strong-coupling
approach based upon the $S$ matrix, to give a justification of the backflow
terms; in Sec.~\ref{sec:backflow}, we describe the backflow correlations; 
in Sec.~\ref{sec:MIT}, we present some results for the metal-insulator 
transition and the insulating phase; finally, in Sec.~\ref{sec:conc}, we draw 
our conclusions.

\section{Model and variational wave function}\label{sec:model}

We consider the Hubbard model with extended hopping in one-dimensional (1D) and
two-dimensional (2D) square lattices.
\begin{equation}\label{eq:hubbard}
{\cal H} =  -\sum_{i,j,\sigma} t_{ij}
c^{\dagger}_{i,\sigma} c^{\phantom{\dagger}}_{j,\sigma} +
U \sum_{i} n_{i,\uparrow} n_{i,\downarrow},
\end{equation}
where $c^{\dagger}_{i,\sigma}$ ($c^{\phantom{\dagger}}_{i,\sigma}$) denotes 
the creation (annihilation) operator of one electron on site $i$ with spin 
$\sigma=\uparrow,\downarrow$,
$n_{i,\sigma}=c^{\dagger}_{i,\sigma}c^{\phantom{\dagger}}_{i,\sigma}$
is the electron density, $t_{ij}$ the hopping parameters, and $U$ the on-site
Coulomb repulsion. In the following, we will consider nearest- and 
next-nearest-neighbor hoppings that will be respectively denoted by $t$ and 
$t^\prime$ in 2D and by $t_1$ and $t_2$ in 1D. We will focus our attention on 
the half-filled case with $L$ electrons on $L$ sites. 

Both metallic and insulating phases can be constructed in a variational
approach. In a first step, one constructs uncorrelated wave functions given by
the ground state $|\rm{BCS}\rangle$ of a superconducting 
Bardeen-Cooper-Schrieffer (BCS) Hamiltonian,~\cite{grosbcs,zhang88}
\begin{equation}\label{eq:meanfield}
{\cal H}_{\rm{BCS}} = \sum_{q,\sigma} \epsilon_q
c^{\dagger}_{q,\sigma} c^{\phantom{\dagger}}_{q,\sigma}
+ \sum_{q} \Delta_q
c^{\dagger}_{q,\uparrow} c^{\dagger}_{-q,\downarrow} + \rm{h.c.},
\end{equation}
where both the free-band dispersion $\epsilon_q$ and the pairing amplitude 
$\Delta_q=\Delta_{-q}$ are variational functions. We use the parametrization
\begin{eqnarray}
\epsilon_q &=& -2\tilde{t} [\cos(q_x)+\cos(q_y)] -4\tilde{t}^\prime \cos(q_x)\cos(q_y) -\mu \nonumber \\
\Delta_q   &=& \Delta_1 [\cos(q_x)-\cos(q_y)], \nonumber 
\label{eq:def_epsilon_delta1}
\end{eqnarray}
on the 2D square lattice and the parametrization 
\begin{eqnarray}
\epsilon_q &=& -2\tilde{t}_1 \cos(q) -2\tilde{t}_2 \cos(2q) -\mu \nonumber \\
\Delta_q   &=& \Delta_1 \cos(q)+\Delta_2 \cos(2q) +\Delta_3 \cos(3q), \nonumber 
\label{eq:def_epsilon_delta2}
\end{eqnarray}
on the 1D lattice. The next-nearest-neighbor hopping parameters, as well as 
the effective chemical potential $\mu$ and the pairing fields, are variational
parameters to be optimized. $\tilde{t}=\tilde{t}_1=1$ is kept fixed to set the
energy scale. The correlated state $|\Psi\rangle$, without backflow, is then 
given by
\begin{equation}\label{eq:eq_BCS}
|\Psi\rangle = {\cal J} |\textrm{BCS}\rangle,
\end{equation}
where 
\begin{equation}\label{eq:jastrow}
{\cal J}=\exp(-\frac{1}{2} \sum_{i,j} v_{i,j} n_i n_j),
\end{equation} 
is a density-density Jastrow factor (including the on-site Gutzwiller term 
$v_{i,i}$), with the $v_{i,j}$ being optimized independently for every distance
$|i-j|$. Notably, within this kind of wave function, it is possible to obtain 
a pure (i.e., non-magnetic) Mott insulator by considering a sufficiently 
strong Jastrow factor,~\cite{capello} i.e., $v_q \sim 1/q^2$ ($v_q$ being the 
Fourier transform of $v_{i,j}$) and a metallic state whenever $v_{q}\sim 1/q$.
In fact, it has been shown that $v_q \sim 1/q^2$ is necessary to have a 
vanishing quasi-particle weight in fermionic models,~\cite{capello2} or to
have a vanishing condensate fraction in bosonic models.~\cite{capello3}
Nonetheless, much more difficult is to demonstrate by numerical simulations
that such a Jastrow term is sufficient to obtain exponential correlations, 
which are suitable for a fully gapped Mott insulator.

As recently shown,~\cite{tocchio1} the projected BCS state $|\Psi\rangle$ can 
be rather poor for large on-site interactions, in presence of frustration. 
In order to overcome this limitation, a further improvement of the variational
wave function is needed. One possibility (which is discussed in the next 
section) is to consider the strong-coupling approach, based upon the $S$-matrix 
expansion.~\cite{macdonald,gros87} Unfortunately, this method cannot be 
easily handled by quantum Monte Carlo simulations on large systems. However,
an alternative, but somewhat related, approach may be defined in terms of
backflow correlations, which is stable and accurate even for large system 
sizes.

\section{The large-$U$ limit and the $S$-matrix expansion}\label{sec:Smatrix}

Let us consider the limit of $U/t \to \infty$. Here, there are no doublons 
and the wave function can be written as 
$|\Psi_{\infty}\rangle={\cal P}_G |\textrm{MF}\rangle$, where the full 
Gutzwiller projector ${\cal P}_G=\prod_{i} (1-n_{i,\uparrow}n_{i,\downarrow})$
removes all double occupancies, while $|\textrm{MF}\rangle$ denotes an 
uncorrelated mean-field state, such as the $|\textrm{BCS}\rangle$ state 
introduced in the previous section. Then, the strong-coupling approach allows 
one to define and construct a variational wave function for the Hubbard model 
at {\it large but finite} values of $U/t$ by
\begin{equation}\label{eq:S_matrix}
|\Psi_S\rangle = e^{-iS} |\Psi_{\infty}\rangle,
\end{equation}
where the operator $S$ can be determined by using a 
recursive scheme. At the lowest order in $t/U$, 
$iS=(T^{+}-T^{-})/U$, where $T^{+}$ and $T^{-}$ 
correspond to the kinetic terms that increase and 
decrease the number of doubly
occupied sites by one:~\cite{macdonald,gros87}
\begin{eqnarray}
T^{+} &=& - \sum_{i,j,\sigma} t_{ij} n_{i,-\sigma}
c^{\dagger}_{i,\sigma}c^{\phantom{\dagger}}_{j,\sigma} (1-n_{j,-\sigma}) \\
T^{-} &=& - \sum_{i,j,\sigma} t_{ij} (1-n_{i,-\sigma})
c^{\dagger}_{i,\sigma}c^{\phantom{\dagger}}_{j,\sigma} n_{j,-\sigma}.
\end{eqnarray}
However, the wave function of Eq.~(\ref{eq:S_matrix}) is hard to handle (both
analytically and numerically), since $S$ is non-diagonal in the basis where the 
electrons have defined positions in the lattice (in contrast to the Jastrow
term, which is diagonal). In the large-$U$ limit, one can further expand the 
exponential and obtain:
\begin{equation}\label{eq:S_exp}
|\Psi_S\rangle \simeq (1-iS)|\Psi_{\infty}\rangle.
\end{equation}
This state has non-vanishing weights only for electronic configurations with
zero or one doublon. Indeed, $|\Psi_{\infty}\rangle$ does not contain doubly 
occupied sites and $S$ may generate at most one doublon. Therefore, after the
expansion of the exponent, the wave function is no longer size consistent. 
Let us see what can be learned from this strong-coupling approach. Suppose 
that an electronic configuration $|x_0\rangle$ has no doubly occupied sites, 
then, since ${\cal P}_G|x_0\rangle=|x_0\rangle$, from Eq.~(\ref{eq:S_exp}) 
we have that:
\begin{equation}\label{eq:zerodoublon}
\langle x_0|\Psi_S\rangle = \langle x_0|\textrm{MF}\rangle;
\end{equation}
on the other hand, if the configuration $|x_1\rangle$ contains one doublon 
(with the holon in one of its neighboring sites), we have that:
\begin{equation}\label{eq:onedoublon}
\langle x_1|\Psi_S\rangle = \pm \frac{t}{U} \langle y_0|\textrm{MF}\rangle
\pm \frac{t}{U} \langle z_0|\textrm{MF}\rangle,
\end{equation}
where $|y_0\rangle$ and $|z_0\rangle$ are the (only) two possible 
configurations with no doubly occupied sites that are connected to 
$|x_1\rangle$ by $S$; signs depend upon the convention used for labeling 
electronic configurations. Therefore, the weight of a configuration with one 
doublon is related to the weight of two configurations without doublons
(times $t/U$).

This procedure for defining a variational wave function is exact for $U \gg t$,
when at most {\it one} doublon is present in the electronic configurations;
however, one is mostly interested into the case where an arbitrary number of 
doublons and holons are present (relevant for the generic case with 
$U \sim t$), and the straightforward generalization of this formalism becomes
much more elaborated and difficult to implement in numerical calculations.
 
\section{Backflow correlations}\label{sec:backflow}

\subsection{Definition of backflow terms}

As introduced in Ref.~\onlinecite{tocchio1}, backflow correlations modify the 
single-particle eigenstates $\phi_k(\boldsymbol{r}_{i,\sigma})$ of the 
mean-field Hamiltonian, like for example ${\cal H}_{\textrm{BCS}}$ defined in 
Eq.~(\ref{eq:meanfield}), according to the electronic configuration on the 
lattice:
\begin{eqnarray}\label{eq:backflow}
\phi_{k}^{b}(\boldsymbol{r}_{i,\sigma}) & \equiv & \tilde{\epsilon}
\phi_{k}(\boldsymbol{r}_{i,\sigma})+ \eta_1 \sum_{j \textrm{n.n.} i}
D_iH_j \phi_{k}(\boldsymbol{r}_{j,\sigma}) \nonumber \\
& + & \eta_2 \sum_{j \textrm{n.n.n.} i}
D_i H_j \phi_{k}(\boldsymbol{r}_{j,\sigma}),
\end{eqnarray}
where $\tilde{\epsilon}=\epsilon$ if the site $i$ is doubly occupied and is 
surrounded by at least one empty site, while $\tilde{\epsilon}=1$ in all the
other cases. Here $\epsilon, \eta_l,(l=1,2)$ are variational parameters to be 
optimized, $D_i=n_{i,\uparrow}n_{i,\downarrow}$ and 
$H_{i}=h_{i,\uparrow}h_{i,\downarrow}$, with $h_{i,\sigma}=1-n_{i,\sigma}$.
Moreover, the shorthand notations n.n. and n.n.n. indicate nearest- and
next-nearest-neighbor sites, respectively. In this way, already the determinant
part of the wave function includes correlation effects, strongly improving 
the accuracy of the many-body state. When backflow correlations are included
in the Slater determinant, the state will be denoted by $|\Psi^b\rangle$. 
This is a substantial improvement with respect to Jastrow factors, where 
electron-electron correlation is included only via a multiplicative term 
(i.e., giving a ``classical'' potential acting on the electronic 
configuration). 

The  results for the $S$-matrix expansion may be compared with the ones of 
backflow correlations for large electron-electron interactions, namely when
at most one doublon is present. Indeed, whenever no doubly-occupied sites
are present, we have that 
\begin{equation}
\langle x_0|\Psi^b\rangle = \langle x_0|\textrm{MF}\rangle
\end{equation}
(where we dropped the Jastrow weight, since all sites are singly occupied, 
and it represents a multiplicative constant). We discuss now the case of a 
single doublon and note that the backflow corrections~(\ref{eq:backflow}) 
are valid for both up- and down-spin orbitals. We consider here, for 
simplicity, only nearest-neighbor backflow terms, with parameters $\epsilon$ 
and $\eta=\eta_1$, and obtain: 
\begin{eqnarray}\label{eq:back_largeU}
\langle x_1|\Psi^{b}\rangle =&& {\cal J}(x_1)
\left[ \pm \eta \epsilon \langle y_0|\textrm{MF}\rangle 
\pm \eta \epsilon \langle z_0|\textrm{MF}\rangle \right . \nonumber \\
+&& \left . \epsilon^2 \langle x_1|\textrm{MF}\rangle 
+ \eta^2 \langle {\tilde x_1}|\textrm{MF}\rangle \right],
\end{eqnarray}
where, in analogy with the $S$-matrix expansion, backflow correlations act on 
configurations with doubly occupied sites (i.e., $|x_1\rangle$), transferring 
some weight to configurations without doubly-occupied sites ($|y_0\rangle$ 
and $|z_0\rangle$ are the same two configurations that appear in the 
strong-coupling expansion); in addition there is some weight also coming from 
configurations with one holon-doublon couple (i.e., the original
configuration $|x_0\rangle$ and a configuration where the two electrons of the
doubly-occupied site have been transferred to the empty site 
$|{\tilde x}_1\rangle$). The signs of the second and third terms of 
Eq.~(\ref{eq:back_largeU}) are the same as the ones appearing in 
Eq.~(\ref{eq:onedoublon}). The relative weights of the admixture in 
Eq.~(\ref{eq:back_largeU}) is controlled by the variational parameters $\eta$
and $\epsilon$. 

Therefore, our way to define the backflow correlations is strictly related to 
the strong-coupling approach; moreover, it allows us to consider many couples 
of holons and doublons, with a moderate computational cost, since they just
re-define the Slater determinant of the wave function. Finally, we would like 
to mention that the backflow wave function, on the contrary to the 
strong-coupling approach of Eqs.~(\ref{eq:S_exp}), (\ref{eq:zerodoublon}),
and~(\ref{eq:onedoublon}), is size consistent. 

The terms in Eq.~(\ref{eq:backflow}) are the dominant contributions to the 
backflow operator, having a qualitative influence on the properties of the 
correlated wave functions. In addition, we can also take into account further 
terms that are useful in the intermediate-coupling regime and correspond
to all possible hopping processes:
\begin{eqnarray}\label{eq:backflow2}
&& \phi_{k}^{b}(\boldsymbol{r}_{i,\sigma}) \equiv \tilde{\epsilon}_i
\phi_{k}(\boldsymbol{r}_{i,\sigma})+ \eta_1 \sum_{j \textrm{n.n.} i}
D_i H_j \phi_{k}(\boldsymbol{r}_{j,\sigma}) + \nonumber \\
&& \eta_2 \sum_{j \textrm{n.n.n.} i}
D_i H_j \phi_{k}(\boldsymbol{r}_{j,\sigma}) + \nonumber \\
&& \beta_1 \sum_{j \textrm{n.n.} i}
n_{i,\sigma}h_{i,-\sigma}n_{j,-\sigma}h_{j,\sigma}
\phi_{k}(\boldsymbol{r}_{j,\sigma})  + \nonumber \\
&& \beta_2 \sum_{j \textrm{n.n.n.} i}
n_{i,\sigma}h_{i,-\sigma}n_{j,-\sigma}h_{j,\sigma}
\phi_{k}(\boldsymbol{r}_{j,\sigma}) + \nonumber \\ 
&& \gamma_1 \sum_{j \textrm{n.n.} i}
\left( D_i n_{j,-\sigma}h_{j,\sigma} + n_{i,\sigma} h_{i,-\sigma}H_{j}\right)
\phi_{k}(\boldsymbol{r}_{j,\sigma}) + \nonumber \\
&& \gamma_2 \sum_{j \textrm{n.n.n.} i}
\left( D_i n_{j,-\sigma}h_{j,\sigma} + n_{i,\sigma} h_{i,-\sigma}H_{j}\right)
\phi_{k}(\boldsymbol{r}_{j,\sigma}),
\end{eqnarray}
where, in addition to $\epsilon$ and $\eta_l$, $\beta_l,\gamma_l (l=1,2)$ are 
variational parameters to be optimized. All results presented here are obtained
by fully incorporating the backflow corrections and optimizing 
individually~\cite{sorella} every variational parameter in $\epsilon_q$ and 
$\Delta_q$ of Eq.~(\ref{eq:meanfield}), in the Jastrow factor $\cal J$ of 
Eq.~(\ref{eq:jastrow}), as well as for the backflow corrections.

\begin{figure}
\includegraphics[width=\columnwidth]{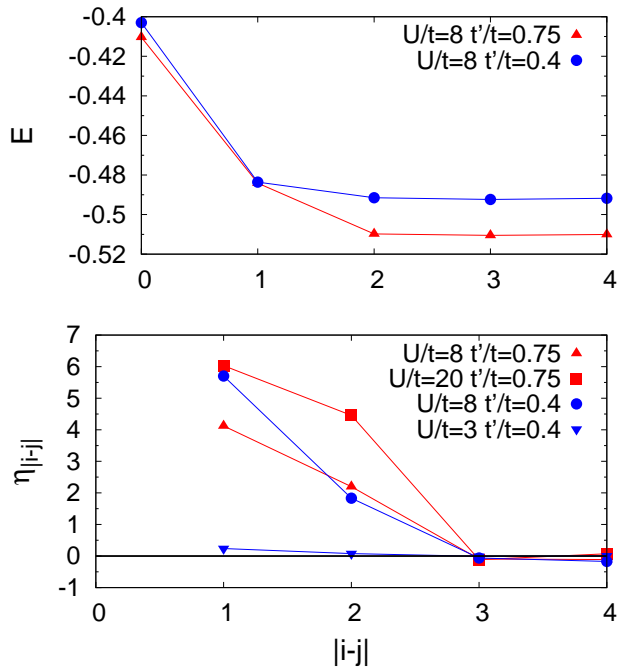}
\caption{\label{fig:back_est}
(Color online) Upper panel: Ground-state energy as a function of the range of 
backflow parameters on the 2D square lattice for $t^\prime/t=0.4$ and $0.75$ 
at $U/t=8$. Range equal to zero corresponds to no backflow correlations in the
wave function. Lower panel: Optimized backflow parameters $\eta_{|i-j|}$ as a 
function of the distance $|i-j|$. Results for $U/t=3$ and $t^\prime/t=0.4$ 
correspond to a metallic state, the others to insulating states. Data are shown
for a lattice with $L=98$ sites.} 
\end{figure}

\begin{figure}
\includegraphics[width=\columnwidth]{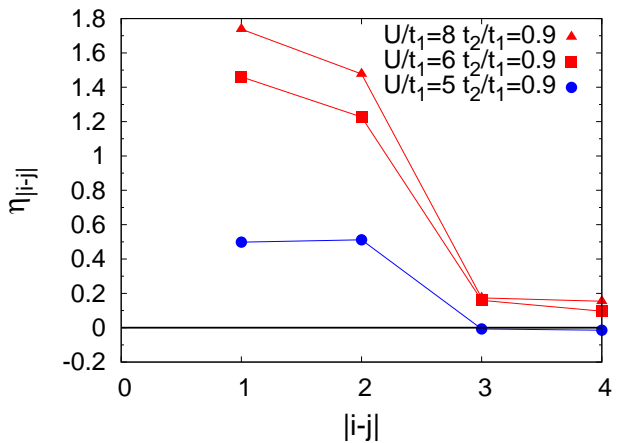}
\caption{\label{fig:back_est_1D}
(Color online)  Optimized backflow parameters $\eta_{|i-j|}$ as a function of 
the distance $|i-j|$ for the 1D lattice. The point at $U/t_1=5$ lies in the 
metallic phase, while the other points are located in the insulating region
of the phase diagram. Data are shown for a lattice with $L=120$ sites.} 
\end{figure}

\begin{figure}
\includegraphics[width=\columnwidth]{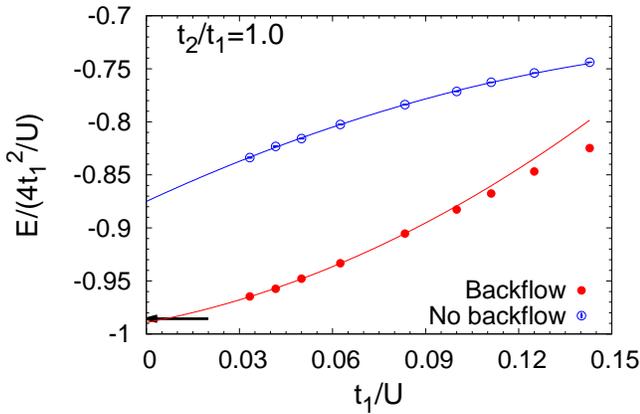}
\caption{\label{fig:estrap}
(Color online) The energy (in unit of $4t_1^2/U$), as a function of $t_1/U$,
for the 1D Hubbard model with $t_2/t_1=1$. Filled (empty) circles denote the 
results with (without) backflow correlations. The DMRG energy of the 
corresponding Heisenberg model with $J_2/J_1=1$ is shown by an 
arrow.~\cite{chitra} Data are presented for a $L=120$ lattice size.} 
\end{figure}

\subsection{Long-range correlations}

We consider now the extension of the backflow parameters to further distances.
In practice, we allow in Eqs.~(\ref{eq:backflow}) and~(\ref{eq:backflow2})
different backflow parameters $\eta_{|i-j|}, \beta_{|i-j|}, \gamma_{|i-j|}$
for every distance of the lattice. We present the results in 
Fig.~\ref{fig:back_est} for the frustrated 2D square lattice. Here, it is 
clear that all backflow parameters are irrelevant for all distances larger 
than few lattice spacing; in particular, they are vanishing for all distances 
not connected by the hopping amplitude. Most importantly, a remarkable gain 
in energy is obtained only by a suitable optimization of the backflow 
parameters up to second neighbors, while poor results are obtained when there 
is no backflow (or its range is smaller than the one of the hopping). 
These results suggest that backflow correlations are indeed necessary to 
successfully describe the effect of frustrating couplings. Furthermore, 
we show in Fig.~\ref{fig:back_est_1D} the behavior of $\eta_{|i-j|}$ for the 
1D lattice up to the fourth distance. Also in this case, the optimized 
backflow parameters become much smaller for distances not connected by the 
hopping amplitude.  

The effect of backflow correlations in the large $U$ limit is presented in 
Fig.~\ref{fig:estrap}, where energies for large $U/t_1$ are shown for the 
frustrated 1D lattice with $t_2/t_1=1$. Only the presence of backflow 
correlations in the wave function allows for a proper extrapolation 
to the Heisenberg limit. Here, the energy of the Heisenberg model is provided 
by density-matrix renormalization group (DMRG) calculations,~\cite{chitra} 
which are numerically exact for 1D models. 

In order to assess the accuracy of backflow correlations, we compare the 
energies obtained within the variational Monte Carlo (VMC) approach of 
Sec.~\ref{sec:model} with the data obtained by means of Green's Function Monte
Carlo (GFMC) within the fixed-node approximation.~\cite{gfmc} This approach 
systematically improves the variational wave function, extracting the best 
ground-state approximation with the same nodes of the starting variational 
ansatz. The data presented in Table~\ref{tab:energyI}, for the insulating 
phase of the frustrated square lattice, show that the variational energy, 
in presence of backflow correlations, is lower than the GFMC energy, when the
starting variational ansatz does not include the backflow term. This means 
that backflow correlations are necessary in order to properly describe 
the ground-state wave function of the frustrated model, capturing its correct 
signs. In Table~\ref{tab:energyII}, we show also data for the unfrustrated 
square lattice with $t^\prime=0$. Here, backflow correlations still strongly 
improve the accuracy of the variational ansatz, but are less crucial in 
reproducing the correct signs of the ground-state wave functions. Indeed, 
for $t^\prime=0$, the signs of the $|\Psi\rangle={\cal J}|\textrm{BCS}\rangle$ 
wave function become exact in the limit $U/t\to \infty$ (e.g., the GFMC results
with or without backflow correlations are very close for large $U/t$).
 
\begin{table}
\caption{\label{tab:energyI} 
Ground-state energies for the Hubbard model on the square lattice with 
$t^\prime/t=0.75$ and $L=162$, obtained by means of variational (VMC) and 
Green's Function Monte Carlo (GFMC) methods. Data are shown both in presence 
of and without backflow correlations.}
\begin{tabular}{ccc}
\hline
$U/t$ & VMC NO backflow & GFMC NO backflow \\
\hline \hline
8     & -0.4048(1)      & -0.5072(3) \\
10    & -0.3409(1)      & -0.4223(3) \\
12    & -0.2970(1)      & -0.3617(2) \\
16    & -0.2373(1)      & -0.2816(2) \\
\hline 
$U/t$ & VMC backflow    & GFMC backflow \\
\hline \hline
8     & -0.5073(1)      & -0.5307(2) \\
10    & -0.4282(1)      & -0.4455(2) \\
12    & -0.3704(1)      & -0.3831(2) \\
16    & -0.2901(1)      & -0.2976(1) \\
\hline \hline
\end{tabular}
\end{table}

 \begin{table}
\caption{\label{tab:energyII} The same as in Table~\ref{tab:energyI}, but
with $t^\prime=0$.}
\begin{tabular}{ccc}
\hline
$U/t$ & VMC NO backflow & GFMC NO backflow \\
\hline \hline
6     & -0.5307(1)      & -0.6444(1) \\
8     & -0.3955(1)      & -0.5045(2) \\
10    & -0.3353(1)      & -0.4203(3) \\
12    & -0.2930(1)      & -0.3634(3) \\
\hline 
$U/t$ & VMC backflow    & GFMC backflow \\
\hline \hline
6     & -0.5961(1)      & -0.6448(3) \\
8     & -0.4803(1)      & -0.5152(3) \\
10    & -0.4022(1)      & -0.4278(2) \\
12    & -0.3451(1)      & -0.3651(2) \\
\hline \hline
\end{tabular}
\end{table}

\subsection{Effect of backflow terms on the Jastrow factor}

\begin{figure}
\includegraphics[width=\columnwidth]{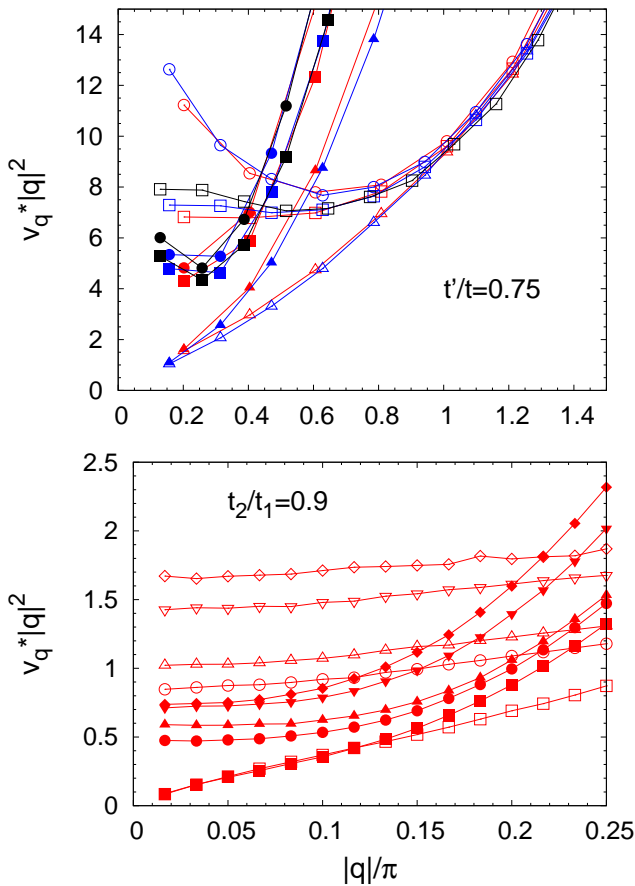}
\caption{\label{fig:jastrow}
(Color online) Upper panel: Fourier transform of the optimized Jastrow factor 
$v_q$ multiplied by $|q|^2$ as a function of $|q|$, along the $(1,1)$ direction
of the Brillouin zone of the square lattice, at $t^\prime/t=0.75$ for
$U/t=7$ (triangles), $U/t=8$ (squares) and $U/t=10$ (circles). Full (empty) 
symbols refer to the presence (absence) of backflow correlations, while 
different colors correspond to different lattice sizes: red ($L=98$), blue 
($L=162$), black ($L=242$). Lower panel: The same quantity for a 1D lattice
with $t_2/t_1=0.9$. Data are shown for $U/t_1=5.4$ (squares), $U/t_1=5.6$ 
(circles), $U/t_1=6$ (up-triangles), $U/t_1=8$ (down-triangles) and $U/t_1=10$ 
(diamonds) on a $L=120$ lattice size. Full (empty) symbols refer to the 
presence (absence) of backflow correlations.}
\end{figure}

Here, we would like to discuss the effect of the backflow corrections on the
behavior of the Jastrow factors across the metal-insulator transition.
As shown in Fig.~\ref{fig:jastrow}, the metallic phase is characterized by 
$v_q \sim 1/q$, while the insulating phase exhibits a $v_q \sim 1/q^2$ 
behavior. A further logarithmic divergence at $q \to 0$ is expected to occur
in 2D.~\cite{capello2,capello3} More details on the relation between the 
Jastrow factor and the conduction properties may be found in 
Appendix~\ref{sec:App_Jastrow}. The presence of backflow correlations in the 
wave function strongly reduces the strength of the Jastrow factor in the 
insulating phase, even if a $v_q \sim 1/q^2$ behavior is always necessary
in order to induce a metal-insulator transition at a finite $U/t$. Indeed, 
a variational wave function including only a local (soft) Gutzwiller factor 
$P=\exp(-v \sum_i n_{i,\uparrow}n_{i,\downarrow})$ and backflow correlations 
is found always in the metallic phase.

\section{Metal-insulator transition and the insulating phase}\label{sec:MIT}

\begin{figure}
\includegraphics[width=0.4\textwidth]{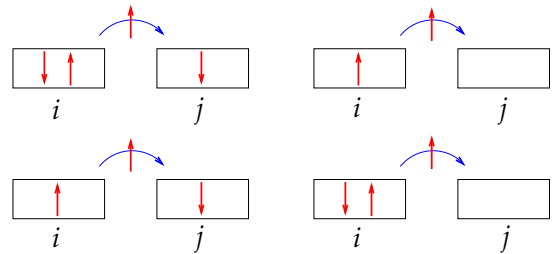}
\caption{\label{fig:hop}
(Color online) Upper raw: Hopping terms of a spin up electron from site $i$ 
to site $j$ that contribute to ${\cal K}_{0}$. Lower raw: hopping terms of a 
spin up electron from site $i$ to site $j$ that contribute to ${\cal K}_{\pm}$.
Lattice sites $i$ and $j$ are represented as boxes and spin up/down electrons 
inside them represent the electronic configurations.} 
\end{figure}

Backflow correlations are a powerful tool to describe the whole insulating 
phase, from the large-$U$ limit down to the metal-insulator transition. 
In the following, we consider the density of doubly occupied sites 
$D=1/L \langle \sum_{i} n_{i,\uparrow}n_{i,\downarrow}\rangle$, 
which is proportional to the interaction energy per site. In addition, we 
split the kinetic energy into two different parts and we compute them 
separately: ${\cal K}_{0}$ is the part where the hopping process does not 
change the total number of holon-doublon couples, while ${\cal K}_{\pm}$ 
denotes the contributions where one holon-doublon couple is created or 
destroyed (see Fig.~\ref{fig:hop}).

\begin{figure}
\includegraphics[width=\columnwidth]{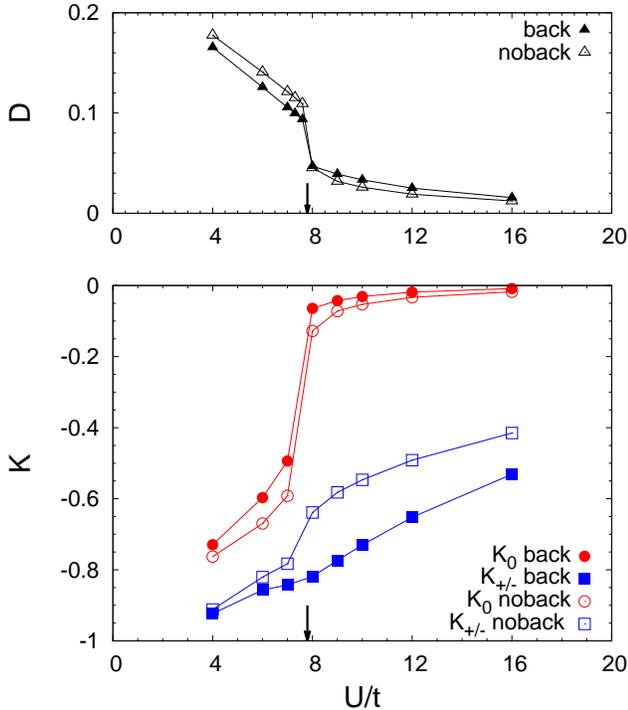}
\caption{\label{fig:kin_square_frust}
(Color online) Upper panel: Density of doubly occupied sites $D$ as a function
of $U/t$. Lower panel: ${\cal K}_{0}$ (circles) and ${\cal K}_{\pm}$ (squares)
as a function of $U/t$. Full (empty) symbols refer to the presence (absence) of 
backflow correlations in the wave function. The metal-insulator transition is 
marked by an arrow. Data refer to a frustrated square lattice with 
$t^\prime/t=0.75$ and $L=162$.} 
\end{figure}

\begin{figure}
\includegraphics[width=\columnwidth]{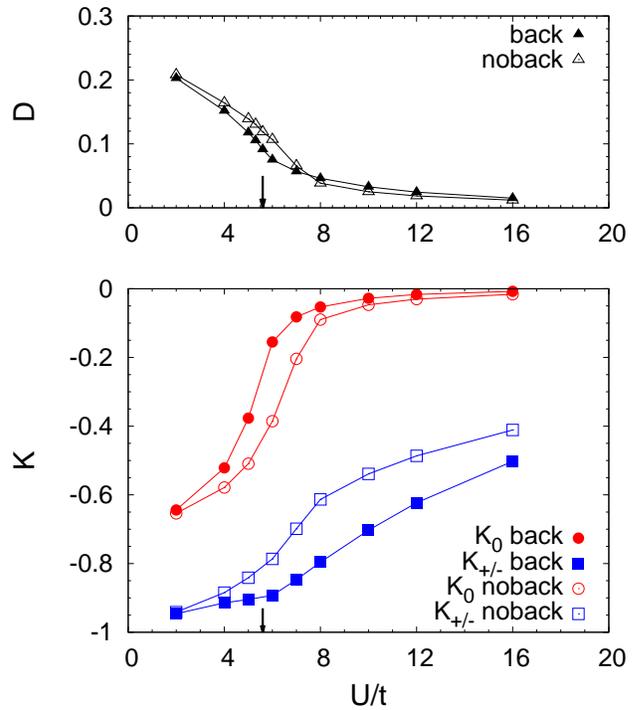}
\caption{\label{fig:kin_square_tpr0}
(Color online) The same as in Fig.~\ref{fig:kin_square_frust}, but for
$t^\prime=0$.}
\end{figure}

\begin{figure}
\includegraphics[width=\columnwidth]{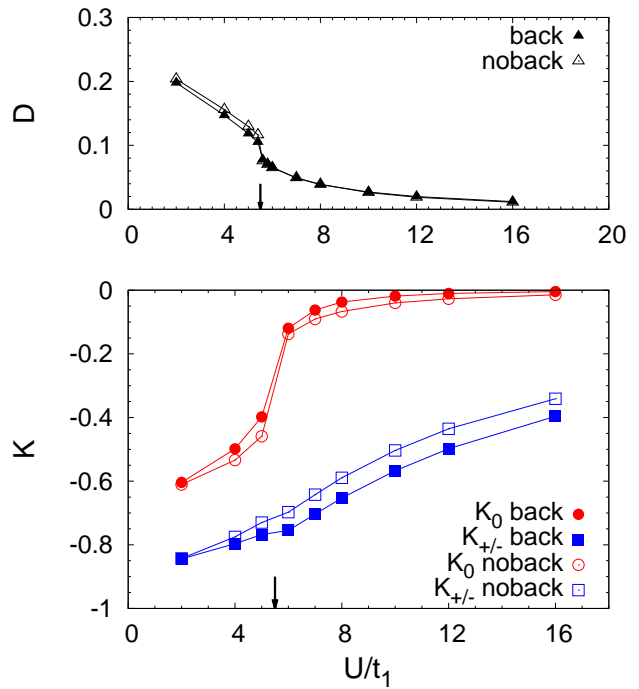}
\caption{\label{fig:kin_1D}
(Color online) The same as in Fig.~\ref{fig:kin_square_frust}, but for a
1D lattice with $t_2/t_1=0.9$ and $L=120$.}
\end{figure}

The results for $D$ and the kinetic terms are presented in 
Figs.~\ref{fig:kin_square_frust},~\ref{fig:kin_square_tpr0} 
and~\ref{fig:kin_1D} as a function of $U$ for the frustrated square lattice 
with $t^\prime/t=0.75$, the unfrustrated one with $t^\prime=0$, and the 1D
lattice with $t_2/t_1=0.9$, respectively. 
The effect of backflow can be clearly seen in the behavior of ${\cal K}_{\pm}$,
which is systematically improved, especially in the insulating phase. This 
trend can be understood by looking at Eq.~(\ref{eq:backflow}): as long as a 
holon-doublon couple is formed, backflow correlations favor its recombination
into single-occupied sites, increasing the part of the kinetic energy that
changes the total number of holon-doublon couples. Moreover, the effect of 
backflow correlations is particularly strong close to the metal-insulator 
transition, where the increasing of the holon-doublon recombination washes out
the jump in ${\cal K}_{\pm}$. On the contrary, the behavior of ${\cal K}_{0}$ 
shows a suppression in presence of backflow correlations, since hopping terms 
that does not change the total number of holon-doublon couples contribute less
to the total energy. We want to point out that, in presence of frustration,
${\cal K}_{0}$ exhibits a clear jump at the metal-insulator transition, 
similarly to the double occupation $D$, both for the square
(see Fig.~\ref{fig:kin_square_frust}) and the 1D lattice 
(see Fig.~\ref{fig:kin_1D}). Instead, for the unfrustrated square lattice, 
the jump vanishes and the metal-insulator transition is characterized only by 
an inflection point (see Fig.~\ref{fig:kin_square_tpr0}). 
 
Finally, we would like to discuss how the variational method allows us to 
estimate the charge gap in the insulating phase, by assessing the
particle-hole excitations over the variational state. This is possible by 
considering just ground-state expectation values, without directly calculating
energy differences. Indeed, within the context of the single-mode
approximation (SMA), which was proposed for the liquid Helium~\cite{feynman2} 
and further applied to fermionic systems,~\cite{girvin,overhauser} it is
possible to find a relation between the particle-hole excitation energy and 
the static structure factor $N(q)=\langle n_{-q}n_q\rangle$ where
$n_q=1/\sqrt{L} \sum_{r,\sigma} \textrm{e}^{iqr}n_{r,\sigma}$ is the 
Fourier transform of the particle density. Indeed, a variational ansatz
of the lowest energy state $|\Psi_q\rangle$, with a given momentum $q$, can be
obtained by applying $n_q$ to the ground state wave function $|\Psi\rangle$ 
(or one approximation for it), namely:
\begin{equation}\label{eq:excit}
|\Psi_q\rangle = n_q|\Psi\rangle.
\end{equation}
The charge gap for the limit $q\to 0$ is then derived in 
Appendix~\ref{sec:App_gap} and results:
\begin{equation}\label{eq:charge_gap}
E_q= -\frac{1}{2\textrm{d}}\left( \lim_{q\to 0} \frac{|q|^2}{N(q)}\right)
\left[ (\Delta r_1)^2 {\cal K}_1 + (\Delta r_2)^2 {\cal K}_2 \right], 
\end{equation}
where d is the dimensionality, ${\cal K}_1$ and ${\cal K}_2$ are the nearest- 
and the next-nearest-neighbor kinetic energy per site, while $\Delta r_1$ and 
$\Delta r_2$ are the distances between nearest- and next-nearest-neighbors 
on the lattice (e.g., $\Delta r_1=1$ and $\Delta r_2=2$ or $\sqrt{2}$ on the 
1D or 2D square lattice).

The metallic phase is characterized by $N(q)\sim q$ for $q \to 0$, which
implies a vanishing gap for particle-hole excitations. On the contrary, in
the insulating phase, $N(q)\sim q^2$ for $q \to 0$, enclosing the fact
that the charge gap is finite. 

\begin{figure}
\includegraphics[width=\columnwidth]{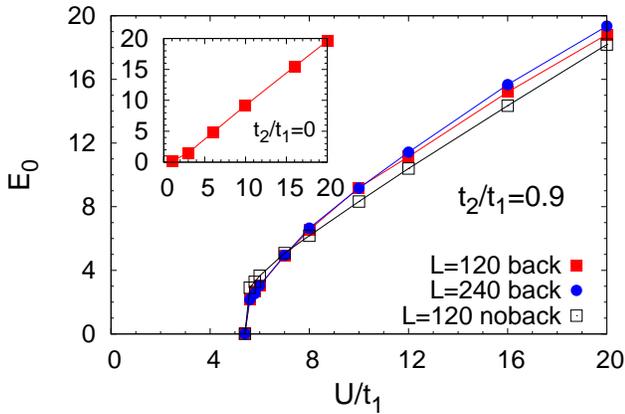}
\caption{\label{fig:charge_gap}
(Color online) Charge gap $E_q$ for $q \to 0$ for the 1D Hubbard model with 
$t_2/t_1=0.9$, on $L=120$ and $L=240$ lattices. Results without backflow 
correlations for the $L=120$ case are shown for comparison. Inset: charge gap 
for the 1D Hubbard model with $t_2=0$ on a $L=120$ lattice size. 
Axis labels are the same of the main plot.} 
\end{figure}

In Fig.~\ref{fig:charge_gap}, we show the gap for $q \to 0$ for the 1D Hubbard
model with $t_2/t_1=0.9$. The gap closes at the metal-insulator transition, 
as expected, and grows up linearly with $U$ for large values of $U/t_1$.
For comparison, we also show the same gap in the case $t_2=0$ (where a Mott
insulating state takes place for infinitesimal values of $U/t_1$). 

\section{Conclusions}\label{sec:conc}

The search for high-quality variational descriptions of the ground-state 
properties of correlated and frustrated electron systems is a long-standing 
issue. The problem lies in the correct description of emergent energy scales 
of order $t^2/U$, which are responsible both for the generation of 
antiferromagnetic correlations and spin-liquid behaviors of frustrated systems,
as well as for incipient and real superconducting state formation.

The description of ``dynamical'' energy scales, viz of energy scales which are 
not explicitly present in the original Hamiltonian but are generated by
dynamical processes (e.g., the super-exchange $J=4t^2/U$), is a challenge for 
both variational approaches and, more generally, for numerical simulations. 
It can be circumvented in the limit of large interactions, i.e., $U\gg t$, 
by using the $S$-matrix expansion outlined in Sec.~\ref{sec:Smatrix}. 
One can then transform the Hubbard model into the $t{-}J$ model, for which 
there is no need to generate the antiferromagnetic energy scale, since it is 
already present explicitly in the model.

The $t{-}J$ model approach has played a key role in developing the RVB approach 
to high-temperature superconductivity. Within this framework, it is however 
not possible to study and correctly describe the Mott-Hubbard transition.
With the backflow corrections, investigated in detail in the present study, 
it is instead now possible to describe, with a high degree of accuracy, the 
dynamical energy-scales of the Hubbard model both in the large- and the 
intermediate-$U$ region. We hence believe that substantial progress has been 
made towards the solution of a long-standing issue, namely the correct 
description of emergent energy scales in correlated and frustrated electron 
systems.

L.F.T. and C.G. acknowledge the support of the German Science Foundation 
through the Transregio 49.

\appendix

\section{Jastrow factors and charge gap}\label{sec:App_Jastrow}

As discussed in Sec.~\ref{sec:MIT}, the insulating or metallic nature of the 
ground state of the Hubbard model can be determined by looking at the static 
structure factor $N(q)=\langle n_{-q}n_q\rangle$. Indeed, the behavior of 
this quantity for $q\to 0$ is related to the presence of a gap in the charge 
excitations of the system by Eq.~(\ref{eq:charge_gap}). The insulating 
phase is characterized by $N(q)\sim q^2$ for $q \to 0$, while in the metallic 
phase $N(q) \sim q$.

A relation between the static structure factor and the Jastrow factor 
${\cal J}=\exp(-1/2 \sum_{q} v_q n_{q} n_{-q})$ has been derived in the 
context of liquid Helium by Reatto and Chester,~\cite{reatto} using a Gaussian
approximation for the probability density associated to the uncorrelated 
ground state wave function $|\textrm{MF}\rangle$. We have that:
\begin{equation}\label{eq:jastrow_q}
N(q)=\frac{N^0(q)}{1+2 v_q N^0(q)},
\end{equation}
where $N^0(q)$ is the static structure factor for the uncorrelated wave 
function $|\textrm{MF}\rangle$. This relation is rigorously valid in the 
weak-coupling regime; nevertheless, a similar behavior is found in the
(strong-coupling) 1D insulating phase.~\cite{capello2} The optimization of 
$v_{i,j}$ for every distance $|i-j|$, will lead, in $q$-space, to 
$v_q \sim 1/q$ for the metallic phase (in any dimension) and to 
$v_q \sim 1/q^2$ (in 1D) and $v_q \sim \log(q)/q^2$ (in 2D), 
see Fig.~\ref{fig:jastrow}. Since $N^0(q)\sim \textrm{const}$ for 
$|\textrm{MF}\rangle=|\textrm{BCS}\rangle$ and $v_q$ diverges for $q \to 0$,
the condition $2v_qN^0(q)\gg 1$ holds in the limit $q \to 0$ and one obtains 
\begin{equation}
N(q) \sim \frac{1}{v_q}.
\end{equation}
Therefore, the small-$q$ behavior of the structure factor reflects the 
small-$q$ behavior of the Jastrow factor. In the 2D insulating phase, although
$v_q \sim \log(q)/q^2$ is found, the static structure factor shows a quadratic 
behavior, i.e., $N(q) \sim q^2$, highlighting the fact that corrections to 
Eq.~(\ref{eq:jastrow_q}) are present in the 2D Mott phase.  

\section{Derivation of the single-particle charge gap}\label{sec:App_gap}

Given the ansatz for the excited state of Eq.~(\ref{eq:excit}), where 
$n_q=1/\sqrt{L} \sum_{r,\sigma} \textrm{e}^{iqr}n_{r,\sigma}$ is the 
Fourier-transformed particle density, the variational estimator of the 
excitation energy is then given by
\begin{eqnarray}
E_q=\frac{\langle \Psi_q|({\cal H}-E_0)|\Psi_q\rangle}{\langle \Psi_q|\Psi_q\rangle}
&=&
\frac{\langle \Psi|n_{-q}[{\cal H},n_q]|\Psi\rangle}{\langle \Psi_q|\Psi_q\rangle} \\ \nonumber
&=&\frac{\langle \Psi|[n_{-q}{\cal H}],n_q|\Psi\rangle}{\langle \Psi_q|\Psi_q\rangle},
\end{eqnarray}
where ${\cal H}$ is the Hamiltonian of Eq.~(\ref{eq:hubbard}).
The sum of both commutators 
$(n_{-q}{\cal H}n_q-n_{-q}n_q{\cal H})+(n_{-q}{\cal H}n_q-{\cal H}n_{-q}n_q)$
is equivalent to the double commutator 
\begin{equation}
[n_{-q},[{\cal H},n_q]]=
n_{-q}({\cal H} n_q-n_q{\cal H})-({\cal H} n_q-n_q{\cal H})n_{-q}
\end{equation}
due to the inversion symmetry $q \leftrightarrow -q$. 
Therefore, the excitation energy is given by
\begin{equation}
E_q=\frac{1}{2}\frac{\langle\Psi| [n_{-q},[{\cal H},n_q]]|\Psi\rangle}{N_q},
\end{equation}
where $N(q)=\langle \Psi|n_{-q}n_q|\Psi \rangle$ is the static structure 
factor for the ground state. The double commutator $[n_{-q},[{\cal H},n_q]]$ 
can be straightforwardly evaluated and involves the kinetic term only, since 
the potential term of the Hamiltonian contains density-density interactions 
that commute with $n_q$:
\begin{equation}
[n_{-q},[{\cal H},n_q]]= \frac{1}{L} \sum_{k,\sigma} 
(\epsilon_{k+q}+\epsilon_{k-q} - 2 \epsilon_{k}) 
c^\dag_{k,\sigma} c^{\phantom{\dagger}}_{k,\sigma}.
\end{equation}
Therefore, in the limit of $q \to 0$, $E_q$ can be written as:
\begin{equation}
E_q= -\frac{1}{2 \textrm{d}}\left( \lim_{q\to 0} \frac{|q|^2}{N(q)}\right)
\left((\Delta r_1)^2 {\cal K}_1 + (\Delta r_2)^2 {\cal K}_2 \right)\, , 
\end{equation}
where d is the dimensionality, ${\cal K}_1$ and ${\cal K}_2$ are the nearest-
and the next-nearest-neighbor kinetic energy per site, while $\Delta r_1$ and 
$\Delta r_2$ are the distances between nearest- and next-nearest-neighbors 
on the lattice (e.g., $\Delta r_1=1$ and $\Delta r_2=2$ or $\sqrt{2}$ on the 
1D or 2D square lattice).


\begin{thebibliography}{99}

\bibitem{feynman} R.P. Feynman and M. Cohen, 
   Phys. Rev. {\bf 102}, 1189 (1956).
\bibitem{ceperley} Y. Kwon, D.M. Ceperley, and R.M. Martin, 
   \prb {\bf 48}, 12037 (1993); \prb {\bf 58}, 6800 (1998).
\bibitem{holzmann} M. Holzmann, D.M. Ceperley, C. Pierleoni, and K. Esler, 
   \pre {\bf 68}, 046707 (2003).
\bibitem{molecules} N.D. Drummond, P. L\'opez R\'ios, A. Ma, J.R. Trail, G.G.
   Spink, M.D. Towler, and R.J. Needs, 
   J. Chem. Phys. {\bf 124}, 224104 (2006); 
   P. L\'opez R\'ios, A. Ma, N.D. Drummond, M.D. Towler, and R.J. Needs, 
   \pre {\bf 74}, 066701 (2006).
\bibitem{tocchio1} L.F. Tocchio, F. Becca, A. Parola, and S. Sorella, 
   \prb {\bf 78}, 041101(R) (2008); 
   see also, F. Becca, L.F. Tocchio, and S. Sorella, 
   J. Phys.: Conf. Ser. {\bf 145}, 012016 (2009).
\bibitem{tocchio2} L.F. Tocchio, A. Parola, C. Gros, and F. Becca, 
   \prb {\bf 80}, 064419 (2009).
\bibitem{tocchio3} L.F. Tocchio, F. Becca, and C. Gros, 
   \prb {\bf 81}, 205109 (2010).
\bibitem{carleo} G. Carleo, S. Moroni, F. Becca, and S. Baroni, 
   arXiv:1007.4260.
\bibitem{fazekas} See for example, P. Fazekas, 
   {\it Lecture Notes on Electron Correlation and Magnetism}, World Scientific
   (1999).
\bibitem{capello} M. Capello, F. Becca, M. Fabrizio, S. Sorella, and E. Tosatti,
   \prl {\bf 94}, 026406 (2005).
\bibitem{yoko} H. Yokoyama and H. Shiba, 
   J. Phys. Soc. Jpn. {\bf 59} 3669, (1990).
\bibitem{macdonald} A.H. MacDonald, S.M. Girvin, and D. Yoshioka, 
   \prb {\bf 37}, 9753 (1988).
\bibitem{gros87} C. Gros, R. Joynt, and T.M. Rice, 
   \prb {\bf 36}, 381 (1987).
\bibitem{baeriswyl} D. Eichenberger and D. Baeriswyl,
   \prb {\bf 76}, 180504 (2007).
\bibitem{grosbcs} C. Gros, 
   \prb {\bf 38}, 931(R) (1988).
\bibitem{zhang88} F.C. Zhang, C. Gros, T.M. Rice, and H. Shiba,
   Supercond. Sci. Technol. {\bf 1}, 36 (1988).
\bibitem{capello2} M. Capello, F. Becca, S. Yunoki, and S. Sorella, 
   \prb {\bf 73}, 245116 (2006).
\bibitem{capello3} M. Capello, F. Becca, M. Fabrizio, and S. Sorella, 
   \prl {\bf 99}, 056402 (2007).
\bibitem{sorella} S. Yunoki and S. Sorella, 
   \prb {\bf 74}, 014408 (2006).
\bibitem{chitra} R. Chitra, S. Pati, H.R. Krishnamurty, D. Sen, 
   and S. Ramasesha, 
   \prb {\bf 52}, 6581 (1995).
\bibitem{gfmc} D.F.B. ten Haaf, H.J.M. van Bemmel, J.M.J. van Leeuwen,
   W. van Saarloos, and D.M. Ceperley, \prb {\bf 51}, 13039 (1995).
\bibitem{feynman2} R.P. Feynman, 
   Phys. Rev. {\bf 94}, 262 (1954).
\bibitem{girvin} S.M. Girvin, A.H. MacDonald, and P.M. Platzman, 
   \prb {\bf 33}, 2481 (1986).
\bibitem{overhauser} A.W. Overhauser, 
   \prb {\bf 3}, 1888 (1971).
\bibitem{reatto} L. Reatto and G.V. Chester, 
   Phys. Rev. {\bf 155}, 88 (1967).

\end{thebibliography}
\end{document}